# Knowledge Management in Software Engineering: A Systematic Review of Studied Concepts, Findings and Research Methods Used[1]

Finn Olav Bjørnson[1] and Torgeir Dingsøyr[1,2]

[1]Norwegian University of Science and Technology Department of Computer and Information Science, Sem Sælandsvei 7-9, 7491 Trondheim, Norway, Tel.:+ 47 73 59 87 16, fax: + 47 73 59 44 66, e-mail: bjornson@idi.ntnu.no

[2]SINTEF Information and Communication Technology, SP Andersens vei 15b, 7465 Trondheim, Norway, Tel.: +47 73 59 29 79, fax: +47 73 59 29 77, e-mail: torgeir.dingsoyr@sintef.no

**Abstract.** *Software engineering is knowledge-intensive work, and how to manage software engineering knowledge has received much attention. This systematic review identifies empirical studies of knowledge management initiatives in software engineering, and discusses the concepts studied, the major findings, and the research methods used. Seven hundred and sixty-two articles were identified, of which 68 were studies in an industry context. Of these, 29 were empirical studies and 39 reports of lessons learned. More than half of the empirical studies were case studies.*

*The majority of empirical studies relate to technocratic and behavioural aspects of knowledge management, while there are few studies relating to economic, spatial and cartographic approaches. A finding reported across multiple papers was the need to not focus exclusively on explicit knowledge, but also consider tacit knowledge. We also describe implications for research and for practice.*

**Keywords:** software engineering, knowledge management, learning software organization, software process improvement, systematic review

# 1. Introduction

Software engineering is a knowledge-intensive activity. For software organisations, the main assets are not manufacturing plants, buildings, and machines, but the knowledge held by the employees. Software engineering has long recognized the need for managing knowledge and the community could learn much from the knowledge-management community, which bases its theories on well-established disciplines such as cognitive science, ergonomics, and management.

As the field of software engineering matures, there is an increased demand for empirically-validated results and not just the testing of technology, which seems to have dominated the field so far. A recent trend in software engineering is an increased focus on evidence-based software engineering, EBSE [43, 67]. Since the volume of research in the field is expanding constantly, it is becoming more and more difficult to evaluate critically and to synthesise the material in any given area. This has lead to an increased interest in systematic reviews (SR) [66] within the field of software engineering.

In this article, we report on a systematic review of empirical studies of knowledge management in software engineering. Our goal is to provide an overview of empirical studies within this field, what kinds of concepts have been explored, what the main findings are, and what research methods are used. More specifically we ask the following research questions:

1. What are the major knowledge management concepts that have been investigated in software engineering?
2. What are the major findings on knowledge management in software engineering?
3. What research methods have been used within the area so far?

Our target readership is three groups that we think will be interested in an overview of empirical research on knowledge management in software engineering: (1) researchers from software engineering who would like to design studies to address important research gaps, and identify relevant research methods; (2) researchers on knowledge management in general, who would be interested in comparing work in the software engineering field to other knowledge-intensive fields; and (3) reflective practitioners in software engineering, who will be interested in knowing what knowledge management initiatives have been made in software companies, or quickly identifying relevant studies, and the major findings and implications from these.

The remainder of this article is structured as follows. Section 2 presents the background and general theories on knowledge management. Section 3 describes the research method that we used to select and review the data material for our research, and presents our chosen framework for analysis. Section 4 presents the results of the systematic review according to our chosen framework. In Section 5, we discuss the findings and their implications. For research, we identify what we belive are the most important research gaps. For practitioners, we provide advice on how to use the results in practice. Section 6 concludes.

## 2. Background

In this chapter, we first give a brief background on knowledge management, then give an overview of theories often referred to in the knowledge management literature. Finally, we give an overview of existing work on knowledge management in software engineering.

### *2.1 Knowledge management*

Knowledge management is a large interdisciplinary field. There is, as a consequence, an ongoing debate as to what constitutes knowledge management. However, it is beyond the scope of this article to engage in that debate. For our purposes, it is sufficient to cite some definitions that are in common use. Davenport has defined *knowledge management* as "a method that simplifies the process of sharing, distributing, creating, capturing and understanding of a company's knowledge" [28]. A related term is *organisational learning*. What does it mean to say that an organisation as a whole learns? According to Stata, this differs from individual learning in two respects [112]: first, it occurs through shared insight, knowledge and shared models; second, it is based not only on the memory of the participants in the organisation, but also on "institutional mechanisms" such as policies, strategies, explicit models and defined processes (we can call this the "culture" of the organisation). These mechanisms may change over time, what we can say is a form of learning.

Knowledge management has received much attention in various fields, which is demonstrated by the publication of two "handbooks" [33, 45], one encyclopaedia [106], and numerous books [25, 28, 109].

Hanssen et al. [55] refer to two main strategies for knowledge management:
- Codification – to systematise and store information that constitutes the knowledge of the company, and to make this available to the people in the company.
- Personalisation – to support the flow of information in a company by having a centralised store of information about knowledge sources, like a "yellow pages" of who knows what in a company.

Earl [44] has further classified work in knowledge management into schools (see Table 1). The schools are broadly categorized as "technocratic", "economic" and "behavioural". The technocratic schools are 1) the systems school, which focuses on technology for knowledge sharing, using knowledge repositories; 2) the cartographic school, which focuses on knowledge maps and creating knowledge directories; and 3) the engineering school, which focuses on processes and knowledge flows in organizations.

The economic school focuses on how knowledge assets relates to income in organizations.

The behavioural school consists of three subschools: 1) the organizational school, which focuses on networks for sharing knowledge; 2) the spatial school, which focuses on how office space can be designed to promote knowledge sharing; and 3) the strategic school, which focuses on how knowledge can be seen as the essence of a company's strategy.

Table 1: Earl's schools of knowledge management.

|       | Technocratic       |                        |                    | Economic         | Behavioural       |                    |                         |
|-------|--------------------|------------------------|--------------------|------------------|-------------------|--------------------|-------------------------|
|       | Systems            | Cartographic           | Engineering        | Commercial       | Organizational    | Spatial            | Strategic               |
| Focus | Technology         | Maps                   | Processes          | Income           | Networks          | Space              | Mindset                 |
| Aim   | Knowledge bases    | Knowledge directories  | Knowledge flows    | Knowledge assets | Knowledge pooling | Knowledge exchange | Knowledge capabilities  |
| Unit  | Domain             | Enterprise             | Activity           | Know-how         | Communities       | Place              | Business                |

There are a number of overview articles of the knowledge management field in the literature. In the following we describe overview articles from management science and information systems.

In the introduction to the book *Challenges and Issues in Knowledge Management* [22], in the field of management consulting, Buono and Poulfelt claim that the field is moving from first to second generation knowledge management. In first generation knowledge management, knowledge was considered a possession, something that could be captured, thus knowledge management was largely a technical issue on how to capture and spread the knowledge through tools like management information systems, data repositories and mechanistic support structures. The second generation of knowledge management is characterized by knowing-in-action. Knowledge is though of as a socially embedded phenomenon, and solutions have to consider complex human systems, communities of practice, knowledge zones, and organic support structures. The change in knowledge management initiatives is seen to go from a planned change approach to a more guided changing approach.

Coming from the field of management consulting, Christensen [26] performed a literature review focusing on special journal issues on knowledge management from 1995-2003. He performed a content analysis of 50 identified papers focusing on knowledge management context, knowledge management outcomes, empirical setting and the key drivers for knowledge management. The finding was that KM writings seem to focus on how to create knowledge and to a lesser degree, how to transfer knowledge. The categories that did not receive adequate coverage were integration, production, measurement, retention and reflection. A second finding was that the drivers for both knowledge creation and knowledge transfer were generic and to a large degree overlapping. He goes on to explore knowledge management in practice through 10 managers from industry and compares his results to the results of the theoretic study. The main conclusion is that KM theory does reflect, in generic terms, the practices that support KM activities, but the challenge is to observe this practical application of generic drivers, which often is difficult to observe in practice.

In the information systems field, Alavi and Leidner [3] summarize literature from different fields, which is relevant to research on knowledge management systems. One of the major challenges in KM according to them is to facilitate the flow of knowledge between individuals so that the maximum amount of transfer occurs. They also conclude that no single or optimal solution to organizational knowledge management can be developed. Instead a variety of approaches and systems needs to be employed to deal with the diversity of knowledge types. Knowledge management is not a monolithic but a dynamic and continuous phenomenon.

Liao gives an overview of technology and applications for knowledge management in a review of the literature from 1995 to 2002 [74]. The review covers knowledge-based systems, data mining, ICT applications, exptert systems, database technology and modelling technology.

Argote et al. [7] conclude a special issue of Management Science with an article that provides a framework for organizing the literature on knowledge management, identifies emerging themes, and suggests directions for further research.

Many have been critical to the concept of knowledge management, and in particular to the use of information technology in knowledge management. Hislop [56] questions the distinction between tacit and explicit knowledge. If explicit knowledge cannot be managed independently, this means that information technology will have a smaller part in knowledge management. This critique is also supported by McDermot, [85] who argues that "if people working in a group don't already share knowledge, don't already have plenty of contact, don't already understand what insights and information will be useful to each other, information technology is not likely to create it". In addition, Swan et al. [114] criticize the knowledge management field for being too occupied with tools and techniques. They claim that researchers tend to overstate the codifiability of knowledge and to overemphasize the utility of IT to give organizational performance improvement. They also warn that "codification of tacit knowledge into formal systems may generate its own pathology: the informal and locally situated practices that allow the firm to cope with uncertainty may become rigidified by the system".

Schultze and Leidner [105] studied discourses of knowledge management in information systems research, and warn that knowledge can be a double-edged sword: too little can result in expensive mistakes, while too much can lead to unwanted accountability. In a study of research on information systems, they found that most existing research is optimistic on the role of knowledge management in organizations, and they urge researchers to give more attention to the critique of knowledge management.

## 2.2 Theories of organizational learning

In cognitive and organization science, we find many models on how knowledge is transferred or learned at an individual and organizational level. We present four theories that are referred to widely: Kolb's model of experiential learning, the double-loop learning theory of Argyris and Schön, Wenger's theory of communities of practice, and Nonaka and Takeuchi´s theory of knowledge creation.

Kolb describes learning from experience ("experiential learning", see [70]) as four different learning modes that we can place in two dimensions. One dimension is how people take hold of experience, with two modes, either relying on symbolic representation – which he calls comprehension, or through "tangible, felt qualities of immediate experience", which he calls apprehension. The other dimension is how people transform experience, with two modes, either through internal reflection, which he refers to as intention, or through "active external manipulation of the external world", which he calls extension.

Kolb argues that people need to take advantage of all four modes of learning to be effective, they "must be able to involve themselves fully, openly, and without bias in new experiences; reflect on and observe these experiences from many perspectives; create concepts that integrate their observations into logically sound theories; and use these theories to make decisions and solve problems" [71].

Argyris and Schön distinguish between what they call single and double-loop learning [9] in organisations. In single-loop learning, one receives feedback in the form of observed effects and then acts on the basis solely of these observations to change and improve the process or causal chain of events that generated them. In double-loop learning, one not only observes the effects of a process or causal chain of events, but also understands the factors that influence the effects [8].

One traditional view of learning is that it is most effective when it takes place in a setting where you isolate and abstract knowledge and then "teach" it to "students" in *POSTPRINT*rooms free of context. Wenger describes this as a view of learning as an individual process where, for example, collaboration is considered a kind of cheating [120]. In his book about communities of practice, he describes a completely different view: learning as a *social phenomenon*. A community of practice develops its own "practices, routines, rituals, artefacts, symbols, conventions, stories and histories". This is often different from what you find in work instructions, manuals and the like. Wenger defines learning in communities of practice as follows:

For individuals: learning takes place in the course of engaging in, and contributing to, a community.
For communities: learning is to refine the practice.
For organisations: learning is to sustain interconnected communities of practice.

Nonaka and Takeuchi [90] claim that knowledge is constantly converted from tacit to explicit and back again as it passes through an organisation. By tacit knowledge [94] we mean knowledge that a human is not able to express explicitly, but is guiding the behaviour of the human. Explicit knowledge is knowledge that we can represent in textual or symbolic form. They say that knowledge can be converted from tacit to tacit, from tacit to explicit, or from explicit to either tacit or explicit knowledge. These modes of conversion are described as follows:

*Socialization* means to transfer tacit knowledge to another person through observation, imitation and practice, what has been referred to as "on the job" training.
*Externalisation* means to go from tacit knowledge to explicit. Explicit knowledge can "take the shapes of metaphors, analogies, concepts, hypotheses or models".
*Internalisation* means to take externalised knowledge and make it into individual tacit knowledge in the form of mental models or technical know-how.
*Combination* means to go from explicit to explicit knowledge, by taking knowledge from different sources such as documents, meetings, telephone conferences, or bulletin boards and aggregating and systematizing it.

According to Nonaka and Takeuchi, knowledge passes through different modes of conversion, which makes the knowledge more refined and spreads it across different layers in an organisation.

## *2.3 Knowledge management in software engineering*

Companies developing information systems have failed to learn effective means for problem solving to such an extent that they have learned to fail, according to an article by Lyytinen and Robey [81]. One suggested mean to overcome this problem is an increased focus on knowledge management.

There are many approaches to how software should be developed, which also affect how knowledge is managed. A main difference between methods here is if they are plan-based or traditional, which rely primarily on managing explicit knowledge, or agile methods, which primarily rely on managing tacit knowledge [88].

In software engineering, there has been much discussion about how to manage knowledge, or foster "learning software organisations". In this context, Feldmann and Althoff have defined a "learning software organisation" as an organisation that has to "create a culture that promotes continuous learning and fosters the exchange of experience" [50]. Dybå places more emphasis on action in his definition: "A software organisation that promotes improved actions through better knowledge and understanding" [41].

In software engineering, reusing life cycle experience, processes and products for software development is often referred to as having an "Experience Factory" [13]. In this framework, experience is collected from software development projects, and are packaged and stored in an *experience base*. By packing, we mean generalising, tailoring, and formalising experience so that it is easy to reuse.

In 1999, the first workshop on "learning software organizations" was organized in conjunction with the SEKE conference. This workshop has been one of the main arenas for empirical studies as well as technological development related to knowledge management in software engineering.

The May 2002 issue of IEEE Software [77] was devoted to knowledge management in software engineering, giving several examples of knowledge management applications in software companies. In 2003, the book "Managing Software Engineering Knowledge" [40] was published, focusing on a range of topics, from identifying why knowledge management is important in software engineering [78], to supporting structures for knowledge management applications in software engineering, to offering practical guidelines for managing knowledge.

However, Edwards notes in an overview chapter in the book on Managing Software Engineering Knowledge [47] that knowledge management in software engineering is somewhat distanced from mainstream knowledge management.

Several PhD thesis have also been published on aspects of knowledge management that are related to software engineering [16, 18, 36, 117].

In addition, a number of overviews of work on knowledge management in software engineering have previously been published. Rus et al. [100] present an overview of knowledge management in software engineering. The review focuses on motivations for knowledge management, approaches to knowledge management, and factors that are important when implementing knowledge management strategies in software companies. Lindvall et al. [80] describe types of software tools that are relevant for knowledge management, including tools for managing documents and content, tools for managing competence, and tools for collaboration. Dingsøyr and Conradi [37] surveyed the literature for studies of knowledge management initiatives in software engineering. They found eight reports on lessons learned, which are formulated with respect to what actions companies took, what the effects of the actions were, what benefits are reported, and what kinds of strategy for managing knowledge were used.

Despite of the previously published overviews of the field, there is still a lack of broad overviews of knowledge management in software engineering. Our motivation for this study was thus, to give a more thorough and broader overview in the form of a systematic review. This study also covers recent work, and assesses the quality of the research in the field.

## 3. Method

The research method used is a systematic review [66], with demands placed on research questions, identification of research, selection process, appraisal, synthesis, and inferences. We now address each of these in turn.

## 3.1 Planning the review

We started by developing a protocol for the systematic review, specifying in advance the process and methods that we would apply. The protocol specified the research questions, the search strategy, criteria for inclusion and exclusion, and method of synthesis.

The aim of the study was to provide an overview of the empirically studied methods for knowledge management in software engineering, answering the research questions listed in Section 1.

## 3.2 Identification of research

A comprehensive, unbiased search is a fundamental factor that distinguishes a systematic review from a traditional review of the literature. Our systematic search started with the identification of keywords and search terms. We used general keywords in the search in order to identify as many relevant papers as possible.

Table 2: Keywords for our search

| Software engineering keywords | Knowledge management keywords |
|---|---|
| <ul><li>software engineering</li><li>software process</li><li>learning software organization</li></ul> | <ul><li>knowledge management</li><li>tacit knowledge</li><li>explicit knowledge</li><li>knowledge creation</li><li>knowledge acquisition</li><li>knowledge sharing</li><li>knowledge retention</li><li>knowledge valuation</li><li>knowledge use</li><li>knowledge application</li><li>knowledge discovery</li><li>knowledge integration</li><li>knowledge Theory</li><li>organization knowledge</li><li>knowledge engineering</li><li>experience transfer</li><li>technology transfer</li></ul> |

All possible permutations of the software engineering and knowledge management concepts were tried in the search conducted. The following electronic bases were those we considered most relevant [42]: ISI Web of Science, Compendex, IEEE Xplore and the ACM Digital Library.

In addition, we identified two arenas that, to our knowledge, are the only ones that pertain specifically to knowledge management in software engineering: the workshop series on Learning Software Organisations (LSO) from 1999 until 2006, and the book Managing Software Engineering Knowledge [10]. We searched all proceedings from the workshop series and included all chapters from the book.

We performed the search in August 2006, which means that publications up to and including the first quarter of 2006 are included, but some studies in the second quarter might not have been indexed in the databases.

The identification process yielded 2102 articles. This formed the basis for the next step in our selection process.

## 3.3 Selection of primary studies

The first step after the articles had been identified was to eliminate duplicate titles, and titles clearly not related to the review. One researcher (the first author) read through the 2102 titles and removed duplicates and those clearly not related to the field of software engineering. This yielded a result of 762 articles.

After this we obtained the abstract of these articles and both authors read through all abstracts, with the following exclusion criterion.
- Exclude if the focus of the paper is clearly not on software engineering
- Exclude if the focus of the paper is clearly not on knowledge management
- Exclude if the method, tool or theory described is not tested in industry

To narrow the search further we also decided to focus on technical and process knowledge (thus, "software engineering knowledge"). Hence, we also used the criterion
- Exclude if the focus of the paper is on domain knowledge

After each researcher had gone through the papers we compared results. Where we disagreed as to whether to keep or remove a paper, we discussed the matter until we reached agreement.

This process reduced the number of articles to 133, and agreement between researchers was 'good' (Kappa value of 0,655).

The full text for all 133 papers was obtained and both researchers read through all the papers with the same criteria for exclusion in mind. The final number of papers selected for the review was 68. The agreement between researchers at this stage was "moderate" (Kappa value: 0,523).

### *3.4 Quality assessment and classification*

We chose to classify the 68 papers identified along two axes. (1) We wanted to examine what kinds of concept had been tested. To aid us with this we chose the framework for classifying strategies for managing knowledge presented by Earl in [44]. Each researcher classified the 68 papers individually according to the framework, before comparing the results. Disagreements were discussed until a consensus was reached on the classification. (2) We also wanted to examine the scientific rigour of the studies. Here we settled on a simpler classification. All studies included so far had results taken from industry. We further assessed the quality of the selected papers by categorising these into empirical studies and lessons learned reports. The criterion for being accepted as an empirical study and not a report of lessons learned was that the article had a section describing the research method and context. Again, each study was classified individually by the two researchers before comparing the results and discussing problem cases in order to reach agreement. After the quality assessment, we had 29 empirical studies and 39 reports of lessons learned.

### *3.5 Synthesis*

For the synthesis, we chose to only use the papers classified as empirical studies in our framework, in order to avoid problems associated with lessons learned reports stemming from their lack of scientific rigor. We extracted concepts covered, main findings and the research method for each article. One researcher (the first author) focused on the studies in the technocratic schools, while the other researcher (the second author) focused on the behavioural schools.

## 4. Results

Using the framework outlined in Section 3.4, we categorized the 29 empirical studies and 39 reports of lessons learned in Table 3. For a complete listing of papers in each category, see the appendix. Within Earl's framework, we found a heavy concentration on the technocratic schools and a fair mention of the behavioural school. We did not find any papers relating to the economic school with our search criterion. Within the technocratic schools, systems and engineering stand out as areas that have received much attention. Within the behavioural schools, organizational and strategic have received the most attention.

Four of the empirical studies did not fit into Earl's framework. These were classified as studies on the impact of knowledge management initiatives and on knowledge management per se. Thus, we ended up with 25 studies classified as empirical within the framework. Of the 39 reports of lessons learned, two belonged to two categories, which is why we ended up with a sum of 41 for the reports of lessons learned in the table.

**Table 3: Categorized articles**

| | Systems | Cartographic | Engineering | Commercial | Organizational | Spatial | Strategic | SUM |
|---|---|---|---|---|---|---|---|---|
| Empirical studies | 6 | 1 | 12 | 0 | 3 | 0 | 3 | 25 |
| % distribution, empirical studies | *24* | *4* | *48* | *0* | *12* | *0* | *12* | *100* |
| Lessons learned reports | 20 | 0 | 9 | 0 | 2 | 1 | 9 | 41 |
| % distribution, lessons learned reports | *49* | *0* | *22* | *0* | *5* | *2* | *22* | *100* |

Looking at the papers by year of publication, presented in Figure 1, we notice an increasing interest in the area from 1999 onwards. We also notice a shift from more papers on lessons learned to empirical papers from 2003 onwards. The apparent decrease in attention in 2006 is due to our covering only the first third of this year, since our search was conducted in August.

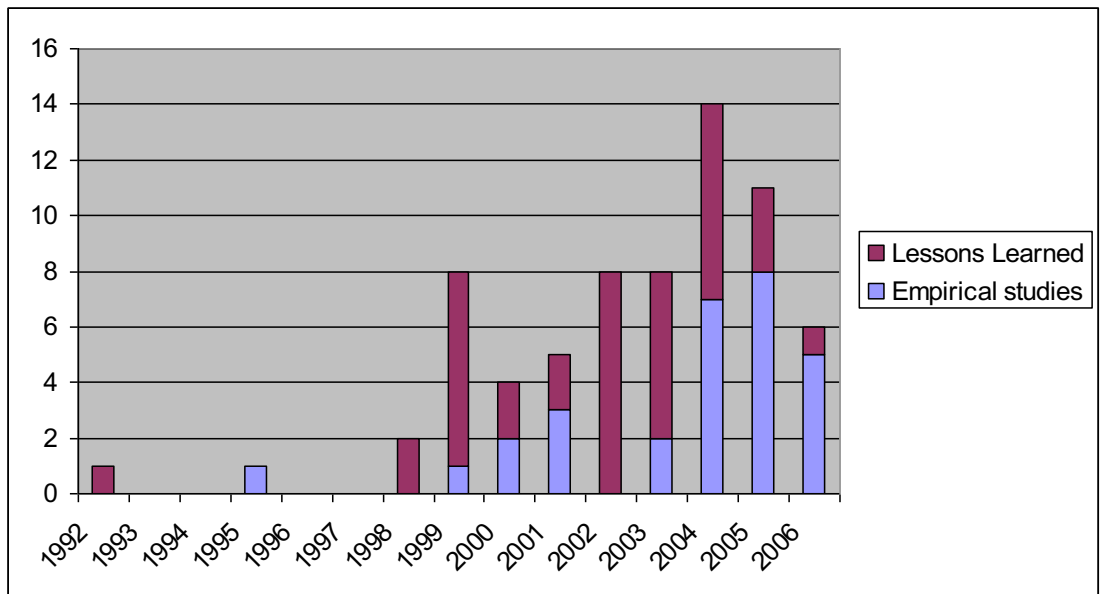

**Figure 1: Publications by year**

To obtain an overview of the research methods used within this field, we used the classification presented in Glass et al. [52]. This was carried out on the 25 papers classified as empirical studies. The result is presented in Table 4. See the appendix for a complete listing of which paper was classified in which category.

**Table 4: Overview of research methods**

|                | Action Research | Case study | Etnography | Experiment | Field study | Sum |
|----------------|-----------------|------------|------------|------------|-------------|-----|
| Systems        | 1               | 3          | 1          |            | 1           | 6   |
| Cartographic   |                 |            | 1          |            |             | 1   |
| Engineering    | 1               | 8          |            | 1          | 2           | 12  |
| Organizational |                 | 3          |            |            |             | 3   |
| Strategic      | 1               |            |            |            | 2           | 3   |
| Sum            | 3               | 14         | 2          | 1          | 5           | 25  |
| %              | 12              | 56         | 8          | 4          | 20          | 100 |

In the following subsections, we present the concepts and main findings from the empirical studies within the main knowledge management schools.

## *4.1 Technocratic schools*

The technocratic schools are based on information or management technologies, which largely support and, to different degrees, condition employees in their everyday tasks. We identified a total of 19 empirical studies and 29 papers on lessons learned in this category. The main focus is on the engineering and systems schools.

## 4.1.1 Systems

As defined by Earl, the systems school is built on the underlying principle that knowledge should be codified in knowledge bases. This is what Hansen et al. refer to as the "codification strategy", and what Nonaka and Takeuchi refer to as externalization.

This school is the longest established school of knowledge management, and it is in this category we found the oldest papers in our search. Most of the papers that were excluded would have been placed in this category, if they had contained empirical results from industry. They could mainly be classified as conceptual analysis and concept implementation, according to Glass's definition. In total, we classified six papers as empirical in this school, and 20 as lessons learned. The empirical papers in this category can broadly be defined as either dealing with the development or use of knowledge repositories. In what follows, we briefly present the major concepts studied in the empirical papers. An overview of concepts and findings can be found in Table 5.

Table 5: Concepts and main findings for the systems school

| School | Concepts | Main Findings | Reference |
|---|---|---|---|
| Systems | Development of knowledge repositories and initial use | Approach to supporting risk in project management | [11] |
| | | Users should be involved in development | [20] |
| | | Approach to support design activities | [24, 111] |
| | Use of knowledge repositories over time | Benefits can be realized quickly, tool remains useful over time, and more benefits accrue over time | [72] |
| | | Tool can be used for different kinds of knowledge than originally intended | [35] |

In [24], Chewar and McCrickard present their conclusions from three case studies investigating the use of their knowledge repository. On the basis of their case studies, they present general guidelines and tradeoffs for developing a knowledge repository. In [20], Bjørnson and Stålhane follow a small consulting company that wanted to introduce an experience repository. On the basis of interviews with the employees, they draw conclusions about attitudes towards the new experience repository, and the content and functionality preferred by the employees. Barros et al. [11] investigate how risk archetypes and scenario models can be used to codify reusable knowledge about project management. They test their approach by an observational analysis in industry. They also describe a feasibility study within an academic environment.

Concerning the actual usage of experience repositories or knowledge bases, Dingsøyr and Røyrvik [35] investigate the practices in a medium-sized software consulting company where knowledge repositories are used in concrete work situations. They found several distinct ways of using the tool and highlight the importance of informal organization and the social integration of the tool in daily work practices. A more formal approach to knowledge management tools is found in [111], where Skuce describes experiences from applying a knowledge management tool in the design of a large commercial software system. Concerning long-term effects of experience repositories, Kurniawati and Jeffrey [72] followed the usage of a combined electronic process guide and experience repository in a small-to-medium-sized software development company for 21 weeks, starting a year after the tool was introduced. They conclude that tangible benefits can be realized quickly and that the tool remains useful with more benefits accruing over time.

### 4.1.2 Cartographic

The principal idea of the cartographic school is to make sure that knowledgeable people in an organization are accessible to each other for advice, consultation, or knowledge exchange. This is often achieved through knowledge directories, or so-called "yellow pages", that can be searched for information as required.

Table 6: Concepts and main findings for the cartographic school

| School | Concepts | Main Findings | Reference |
|---|---|---|---|
| Cartographic | Use of cartographic system | Tool was used for: allocating resources, searching for competence, identifying project oportunities and upgrading skills. | [17] |
| | | Tool enabled learning practice at both individual and company level. | [17] |

We found only one empirical paper within this school and no papers on lessons learned. In [34], Dingsøyr et al. examine a skills management tool at a medium-sized consulting company. They identify four major usages of the tool and point out implications of their findings for future or other existing tools in this category, see Table 6.

## 4.1.3 Engineering

The engineering school of knowledge management is a derivative or outgrowth of business process reengineering. Consequently it focuses on processes. According to our classification, the largest amount of empirical papers came from this school. Two major categories can be identified. The first contains work done by researchers who investigate the entire software process with respect to knowledge management. The second contains work done by researchers who focus more on specific activities and how the process can be improved within this activity. Table 7 gives an overview of concepts and findings for this school.

Table 7: Concepts and main findings for the engineering school

| School | Concepts | Main Findings | Reference |
|---|---|---|---|
| Engineering | Managing knowledge on the software development process | It is feasible to use knowledge management as underlying theory to develop key process areas to supplement the CMM | [14] |
| | | No matter what knowledge management approach you pursue in SPI, you need to create both tacit and explicit knowledge. Tacit is neccesary to change practice, explicit is neccesary to create an organizational memory. | [6] |
| | | A techno-centric approach to SPI may impose unnatural work practices on an organisation and fails to take account of how process improvements might occur spontaneously within a community of practice. | [108] |
| | | The iterative approach of Unified Process ensures large effects in terms of learning, but Unified Process also improves on communication and work distribution in the company. | [51] |
| | | It is possible to define and implement software process in a beneficial and cost-efficient manner in small software organisations. Special considerations must be given to their specific business goals, models, characteristics, and resource | [118] |

| | | limitations. | |
|---|---|---|---|
| | Managing knowledge through formal routines | Formal routines must be supplemented by collaborative, social processes to promote effective dissemination and organizational learning. | [27] |
| | Mapping of knowledge flows | Knowledge mapping can successfully help an organisation to select relevant focus areas for planning future improvement initiatives. | [54] |
| | | Casual maps for risk modeling contributes to organizational learning | [2] |
| | Process for conducting project reviews to extract knowledge | Creating a suitable environment for reflection, dialogue, criticism, and interaction is salient to the conducting of a postmortem. | [32] |
| | | The organizational level can only benefit from the learning of project teams if the knowledge and reasoning behind the process improvements is converted into such an explicit format that it can be utilized for learning in organizational level also. | [101] |
| | Implications of social interaction on knowledge sharing | The focus on the pure codified approach is the critical reason of Tayloristic team failure to effectively share knowledge among all stakeholders of a software project. | [86] |
| | | Increasing the level of reflection in mentor programs can result in more double looped learning. | [19] |

Baskerville and Pries-Heje [15] used knowledge management as the underlying theory to develop a set of key process areas to supplement the Capability Maturity Model (CMM) [93] in a Small and Medium sized Enterprise (SME) software development company. Realising that the CMM did not fit well with an SME company, they helped their case companies to develop new key process areas that focused on managing their knowledge capability. Arent et al. [6] address the challenge of creating organizational knowledge during software process improvement. They argue for the importance of creating organizational knowledge in Software Process Improvement (SPI) efforts and claim that its creation is a major factor for success. On the basis of an examination of several cases, they claim that both explicit and tacit knowledge are required, no matter what approach is pursued. Segal [108] investigates organizational learning in software process improvement. Using a case to initiate and implement a manual of best practice as a basis, she observed that the ideal and actual scenarios of use differed and identified possible reasons for the difference. In [51] Folkestad et al. studied the effect of using the rational unified process as a tool for organizational change. In this case, it was used to introduce development staff to a new technology and methodology. Folkestad et al. concluded that the iterative approach of the unified process had obvious effects on organisational and individual learning. The unified process also resulted in new patterns of communication and a new division of labour being instituted, which had a significant effect on the company. Wangenheim et al. [118] report on their experiences of defining and implementing software processes. They confirm what others have experienced, that it is possible to define and implement software processes in the context of small companies in a beneficial and cost-effective way.

In the papers that focused on specific activities within the process, we identified four major areas: formal routines, mapping of knowledge flows, project reviews, and social interaction. Many of these processes are aimed at stimulating several ways of learning, as, for example, Kolb suggests.

In [27] Conradi and Dybå report on a survey that investigated the utility of formal routines for transferring knowledge and experience. Their main observation was that developers were rather sceptical about using written routines, while quality and technical managers took this for granted. Given this conflict of attitudes, they describe three implications for research on this topic.

Hansen and Kautz [54] argue that if software companies are to survive, it is critical that they improve continuously the services that they provide. Such improvement depends, to a great extent, on the organization's capability to share knowledge and thus on the way knowledge flows in an organization. To investigate knowledge flow, they introduced a tool to map the flows of organisational knowledge in a software development company. Using their new method, they identify potential threats to knowledge flows in an organisation. Also using flow diagrams, Al-Shehab et al. [2] describe how learning from analyses of past projects and from the issues that contributed to their failure is becoming a major stage in the risk management process.

They introduce causal mapping as a method to visualise cause and effect in risk networks. They claim that their method is useful for organisational learning, because it helps people to visualise differences in perceptions.

In [32], Desouza et al. describe two ways of conducting project postmortems. They stress that learning through postmortems must occur at three levels: individual, team, and organization. The paper describes guidelines for when to select different kinds of postmortem, depending on the context and the knowledge that is to be shared. The authors also argue that postmortems must be woven into the fabric of current project management practices. Salo [101] also studies postmortem techniques and concludes that existing techniques lack a systematic approach to validating iteratively the implementation and effectiveness of action taken to improve software processes. Salo studies the implementation of a method to remedy this and observes that the organisational level can only benefit from the learning of project teams if the knowledge and reasoning behind the improvements to processes are converted into an explicit format such that it can be utilized for learning at the organisational level.

In [86], Melnik and Maurer discuss the role of conversation and social interaction effective knowledge sharing in an agile process. Their main finding suggests that the focus on pure codification is the principal reason that Tailoristic teams fail to share knowledge effectively. Moving the focus from codification to socialisation, Bjørnson and Dingsøyr [17] investigated knowledge sharing through a mentor programme in a small software consultancy company. They describe how mentor programmes could be changed to improve the learning in the organization. They also identify several unofficial learning schemes that could be improved.

### *4.2 Behavioural schools*

The behavioural aspects of knowledge management are covered in three schools in Earl's framework: the organizational, spatial, and strategic schools. In our review, we found three empirical studies and two reports of lessons learned in the organizational school, no empirical study and one report of lessons learned in the spatial school, and three empirical studies and nine reports of lessons learned in the strategic school. We present the main concepts and findings from the organizational and strategic schools.

## 4.2.1 Organizational

The organizational school focuses on describing the use of organizational structures (networks) to share or pool knowledge. These structures are often referred to as "knowledge communities". Work on knowledge communities is related to work on communities of practice as described in Section 2.2. An overview of our findings from this school is presented in Table 8.

Table 8: Concepts and main findings for the organizational school

| School | Concepts | Main Findings | Reference |
|---|---|---|---|
| Organizational | How networks are used in software engineering | Networks should be used in addition to other activities when introducing new software engineering methods | [84] |
| | | Description of the role of networks. | [53] |
| | | Networks built on existing informal networks are more likely to be successful | [91] |

The role of networking as an approach to knowledge management has been investigated in three settings where software is developed. Grabher and Ibert [53] discuss what types of network exist in companies, where one case is a software company based in Germany. Mathiassen and Vogelsang [84] discuss how to implement software methods in practice and use two concepts from knowledge management: networks and networking. The network perspective emphasizes the use of technology for sharing knowledge, while networking focuses on trust and collaboration among practitioners involved in software development. The authors stress that knowledge management is highly relevant to understand challenges when introducing new methods for software engineering, and that every company have to find a suitable balance between strategies. In the case company, the emphasis on networks and networking changed considerably during the project. Nörbjerg et al. [91] discuss the advantages and limitations of knowledge networks. They base their discussion on an analysis of two networks related to software process improvement in a medium-sized software company in Europe. Their main finding is that building a network on existing informal networks gave the highest value to the organization.

## 4.2.2 Strategic

In the strategic school, knowledge management is seen as a dimension of competitive strategy. Skandia's views are a prime example [113]. Developing conceptual models of the purpose and nature of intellectual capital has been a central issue. An overview of our findings from this school is presented in Table 9.

Table 9: Concepts and main findings for the strategic school

| School | Concepts | Main Findings | Reference |
|---|---|---|---|
| Strategic | What factors contribute to successful knowledge management | Suggested model, including technological, organizational and human resource-factors | [49] |
| | What learning processes are used in practice | Ongoing interaction between different learning processes important to improve practice | [5] |
| | What strategies exist for managing software engineering knowledge | Found evidence of strategies for codification and personalization in software companies | [116] |

One important issue in the literature on knowledge management has been to identify the factors that lead to the successful management of knowledge. Feher and Gabor [49] developed a model of the factors that support knowledge management. The model includes technological, organizational and human resource factors, and was developed on the basis of data on 72 software development organizations that are contained in the European database for the improvement of software processes.

Another issue of strategic importance is the processes that are in place to facilitate learning. Arent and Nørjeberg [5] analysed three industrial projects for the improvement of software processes, in order to identify the learning processes used. They found that both tacit and explicit knowledge were important for improving practice, and that improvement requires ongoing interaction between different learning processes.

Trittmann [116] distinguish between two types of strategy for managing knowledge: "mechanistic" and "organic". Organic knowledge management pertains to activities that seek to foster innovation, while mechanistic knowledge management aims at using existing knowledge. A survey of 28 software companies in Germany supported the existence of two such strategies. This work parallels the works of Hansen et al. on codification and personalization as important strategies for managing knowledge in the field of management science.

## 4.3 Knowledge management in general

Some studies could not be classified using Earl's framework. These studies can be placed in a broad category that encompasses works that seek to identify the impact of knowledge management initiatives (two empirical studies), and works that investigate knowledge management per se (two empirical studies). An overview of these are presented in Table 10.

Table 10: Concepts and main findings for studies of knowledge management in general.

| School | Concepts | Main Findings | Reference |
|---|---|---|---|
| Knowledge management in general | The impact of knowledge management initiatives | Knowledge pull leads to more effective knowledge management than knowledge push | [1] |
| | | Knowledge needs to be internalized to improve processes | [97] |
| | Factors that enable effective knowledge management | Leadership is the most important enabler for knowledge management | [119] |
| | Factors that contribute to use of knowledge artefacts | Perceived complexity, perceived advantage and perceived risk contribute to the use of knowledge management artefacts | [31] |

### 4.3.1 The impact of knowledge management initiatives

Ajila and Sun [1] investigated two approaches to delivering knowledge to software development projects: "push" and "pull". "Push" means using tools to identify and provide knowledge to potential users. "Pull" means that users themselves have to use repositories and other tools to identify relevant knowledge. On the basis of a survey of 41 software companies in North America, the authors claim that pulling leads to more effective software development.

Ravichandran and Rai [97] studied two models for how the embedding and creation of knowledge influence software process capability. Embedding refers to the process of employing knowledge in standard practices, for example through making work routines, methods and procedures. They found support for a model where knowledge creation has an effect on process capability when the knowledge is embedded after it is created. This means that knowledge has to be internalized before it can be used to improve processes. The study was done as a survey of 103 Fortune 1000 companies and federal and state government agencies in the US.

### 4.3.2 Knowledge management per se

Ward and Aurum [119] describe current practices for managing knowledge in two Australian software companies and explain how leadership, technology, culture, and measurements enable knowledge to be managed effectively and efficiently. They found leadership to be the most significant positive factor for the management of knowledge, but that the tools, techniques, and methodologies that the companies were using were not adequate for managing knowledge effectively.

Desouza et al. [31] examined what factors contribute to the use of knowledge artefacts in a survey of 175 employees in a software engineering organization. They specifically looked at factors that govern the use of explicit knowledge. They found that the following factors relate to the use of explicit knowledge: perceived complexity, perceived relative advantage, and perceived risk.

## 5. Discussion

In this study, we have identified far more studies, particularly empirical studies, than have been reported in previous assessments by Rus et al. [100], Lindvall [80] and Dingsøyr and Conradi [37]. We have shown that although there are not many empirical studies, except for in the systems and engineering schools, there are either empirical studies or reports of lessons learned in all schools except the economic school. Thus, research on knowledge management in software engineering seems to be slowly gaining a broader focus, although research on knowledge management in software engineering is still somewhat distanced from mainstream research on knowledge management.

If we compare the studies found in software engineering to the research directions suggested by Alavi et al. [3], we see that software engineering has primarily addressed the storage and retrieval of knowledge, while topics such as knowledge creation, the transfer and application of knowledge still needs more attention.

We now discuss our findings. We begin with a discussion concerning our first two research questions, then the third, outline implications for research and practice, and end with a discussion of the validity of our study

## *5.1 Major knowledge management concepts and findings*

To answer our two first research questions, we organize the discussion according to Earl's framework, answering "what are the major knowledge management concepts that have been investigated in software engineering?" and "what are the major findings on knowledge management in software engineering?"

In this discussion of what we found, we will also include a discussion of how relevant we think these knowledge management schools are for software engineering. Software engineering is a large field with several disciplines relevant to knowledge management, for example software process improvement. One recent development in software engineering, which has implications for knowledge management activities is whether a company seeks to have agile development processes in place, or rely on tradition development methods such as the waterfall process [88]. Agile software development will focus mainly on knowledge management activities related to tacit knowledge, while the traditional development processes will need activities related to explicit knowledge. In the following, we will discuss the concepts identified in research, and give our opinion on what we think are the most relevant research areas to support agile and traditional software development.

The final selection of papers was divided between the technocratic and behavioural schools, with an emphasis on the technocratic side. This was not surprising, given the general focus of software engineering on the construction of tools and processes. We did not find any examples of what Earl considers economic schools. The reason for this can be twofold, either few software companies track their intellectual capital, or there is little interest in reporting findings from such activities in software engineering.

### **5.1.1 Technocratic schools**

The technocratic schools applied in software engineering can be interesting for other knowledge-intensive disciplines as software engineers are likely to easily adopt new information technology. Looking closer at these schools, we saw a heavy focus on the systems and engineering schools, with barely any mention of the cartographic school. The heavy focus on the systems school can be explained by the software engineering field's focus on implementing new tools [37]. For this school, there is a greater number of lessons learned reports than empirical studies. The main concepts we identified in this school were the development and use of knowledge repositories. There was, however, little to no overlap between the identified papers.

As for findings in this school, there are two studies of the use of knowledge repositories over time, which shows that such tools are actually in use, and have more benefits than the obvious. In Section 2.1, we referred to critique of the codification strategy, and especially a belief that knowledge repositories easily can generate information junkyards. There is not any evidence to support such a claim in software engineering, but we believe there is a heavy publication bias towards success stories. But the cases described in this review shows that it is possible to successfully implement knowledge repositories to work in software companies.

The engineering school is the school that received the most empirical attention, according to our review. Again, we identified two main areas within this school: those focusing on the entire software process and those focusing on particular activities within the process. Within the papers focusing on specific activities, we identified four main areas: formal routines, mapping of knowledge flows, project reviews, and social interaction. As with the systems school, there is little or no overlap between the empirical studies. A possible explanation for the heavy empirical focus within this school is the close fit with work on the improvement of software development processes.

For the findings on whole development process, we see that having an established development process can both improve communication and learning, but we also see that it is important to focus also on sharing tacit knowledge in order to change practice.

In relation to development processes for software, the systems and engineering schools support sharing of explicit knowledge, which is important in traditional software development. Both of these schools require a technical infrastructure in order to facilitate knowledge sharing. However, a finding both from studies in other fields of the systems school [62] and studies of a specific engineering activities, electronic process guides, is that it is difficult to get such technology in actual use [39]. However, many companies have invested in such infrastructure, and this indicates that we need a better understanding of the factors that lead to effective knowledge sharing within these two schools.

That there are so few papers in the cartographic school is interesting. One possible explanation is that the "yellow pages" systems are considered "simple" and undeserving of attention. Earl refers to a number of consulting companies using this school, including McKinsey and Bain (see [55]). However, as the lone study in software engineering shows, such tools have uses other than the obvious, and can stimulate learning both at individual and organizational level. One argument for this school is that although it requires a technical infrastructure, the investment is low because there is no need to codify knowledge. This is a school which is relevant for agile software development, and because of the growing number of such development practices as well as the low cost, we think this is a school which requires further research. A counter-argument could be that tacit knowledge is not as relevant for software development as explicit knowledge, but we see from research on agile development that it is possible to develop high-quality software without making much use of explicit knowledge management [110].

### 5.1.2 Behavioural schools

In the behavioural schools, we found a limited number of papers focusing on organizational and strategic aspects, and no papers focusing on spatial aspects.

The three studies in the organizational school discuss the use of people networks in software organizations. Two of the studies investigated the improvement of software development processes. In Earl's taxonomy, both intra- and interorganizational communities are mentioned as examples. In the software engineering literature, we only find studies made in single organizations. Also, a much debated topic in general knowledge management is what actions management can take in order to support this type of knowledge sharing, what some refer to as knowledge governance. How much should be formal, and what should be left to employees to organize themselves?

As for relevance for software engineering, we believe that this school has the potential to deliver inexpensive solutions for companies, although as the studies in software engineering indicate, there is a debate on whether such initiatives are best left to grow by themselves or if the management should have an active involvement. For software engineering, it could be useful with studies that address this strategy in relation to specific challenges for software development, like challenges with new technology, process improvement or understanding customer needs. This school is relevant for organizations that run multidisciplinary projects, which we believe is the case for most software companies, whether they do agile or traditional development.

As for the spatial school, no empirical studies on software engineering were found in this category. The question is then: Is this something that could be relevant in a software engineering setting? The role of open-plan offices has been studied in other fields, and this is something that also should have an impact on how knowledge is shared in software teams. Many of the agile development methods recommend open-plan offices, and knowing more about what specific effects this has on software development would be valuable.

The empirical studies in the strategic school focus on factors pertaining to successful knowledge management, learning processes, and types of strategy for managing knowledge. It was, perhaps, to be expected that there would not be many articles discussing the strategic importance of knowledge in software engineering supported by empirical findings, because its importance is assumed in most published works on knowledge management in software engineering.

## *5.2 Research methods*

Our third research question addressed research methods used: "What research methods have been used within the area so far?"

Of the 68 studies identified, 39 were reports of lessons learned and 29 were empirical studies. Case studies constituted the largest number of empirical studies (see Table 4), followed by field studies and action research. It is positive that the emphasis on empirical studies has increased (see Figure 1). The apparent dip in 2006 is due to the time at which the search was conducted. We searched the databases in August and most compilers of databases take some months to index their papers; hence, we can only claim to have covered the first third of 2006 fully.

The research methods in the studies that we selected are dominated by case studies, both single and multiple. This is not surprising, considering our limitation on only including studies that performed tests in industry. We found one experiment, and it is not surprising that there are few experiments. Knowledge management is a broad field, and it is difficult to isolate factors for experiments without making the experiment irrelevant.

An important question is then: Is it the right mixture of research methods that are applied to study knowledge management in software engineering? Given the broad nature of knowledge management, we believe it is right to have a large number of case studies. But as the field matures, and we would like to see more studies of the effects of knowledge management, we think we need more in-depth studies in companies, which call for more studies oriented towards ethnography.

Glass et al. [52] found that empirical studies constitute about 5% of published research in software engineering as a whole. Comparing our final findings to the results from our first rough sorting of papers, our final selection constituted about 3% of the initially selected papers. If we assume that Glass's data are representative for the area that we studied within software engineering, we could extrapolate that about 70% of those papers would be conceptual analysis and concept implementation. Most of the papers discarded were indeed conceptual analysis and concept implementation without empirical testing, our results do however, not show a discard number on the empirical criterion as high as 70%. Many studies were also excluded because they were not relevant to either software engineering or knowledge management. Therefore it seems that empirical studies constitute a larger part of the studies on knowledge management in software engineering than in software engineering in general.

### *5.3 Implications for research and practice*
This systematic review has implications both for researchers planning new studies of knowledge management initiatives in software companies, and for practitioners working in software companies who would like to design knowledge management initiatives to meet local needs.

### 5.3.1 Implications for research

For research, we think it is important to have in mind that what kind of knowledge management activities a company should engage in should be determined by how the company develops software. We have distinguished between two types of development which has implications for strategy for knowledge management, namely traditional and agile development.

In this systematic review, we have seen that the knowledge management schools associated with traditional software development so far has received the most attention, namely the systems and engineering schools. This is in line with the observations of Buono and Poulfelt [22], indicating that knowledge management in software engineering is mainly focusing on first generation knowledge management in Section 2.1.

We believe the schools that are relevant to agile software development should be given further attention in the future, as this trend seems to have much influence on industry practice today. Another issue in deciding on priorities for research is the cost of implementing activities in the schools. In general, the schools which do not require codification and a technical infrastructure will be less expensive than the others. Therefore, we argue that in particular the organizational school should be further researched as this school is both relevant for agile and traditional software development, and is inexpensive. Also, the cartographic and spatial schools are good candidates for further research. As for research methods applied, we think there should be a larger focus on in-depth studies, shown through a larger use of ethnographic methods.

### 5.3.2 Implications for practice

As we indicated in implications for research, the technocratic schools are closely related to traditional software development while the behavioural schools are more related to the agile approach to development. The main consideration for practitioners is thus that organisations developing software through a traditional approach will probably benefit more from the technocratic schools, while agile teams would benefit more from behavioural schools.

Practitioners following a traditional approach can find some empirical papers and several lessons learned reports on how to build a knowledge repository. Even though all papers we identified within the systems school are positive it is important to remember the objections to following a pure codification strategy we mentioned in chapter 2.1. We believe there is potential bias in the number of positive reports from this school versus those who report negative results. Our findings from the engineering school also support this view, where several papers underline the importance of not focusing exclusively on codification. An advantage of following the technocratic approach to knowledge management is that there is more material available within this "classical" school. A disadvantage is the cost of implementing strategies relying heavily on codification.

The most important finding from the behavioural schools with implications for practitioners developing in an agile environment would be that network building is more likely to be successful if they are built on already existing networks. Also, the need for diversity in both learning processes and strategies are stressed as important in order to improve practice. An advantage of the behavioural approach to knowledge management is the reduced cost compared to implementing the more application heavy solutions in the technocratic school. However, it has its disadvantage in the relatively few publications on this theme to learn from.

## *5.4 Limitations*

The main threats to validity in this systematic review are threefold: our selection of the studies to be included, correct classification of studies according to Earl's framework of schools in knowledge management, and potential author bias.

As for the selection of studies, only one researcher read through and discarded the first results on the basis of the papers' titles. However, in cases where there was doubt, the papers were included in the next stage. The second and third selection stages, which were based on abstracts and full papers, were carried out by both researchers and we observed a 'good' degree of consensus. In cases where there was disagreement, the issue was discussed until consensus was reached.

Concerning the classification of studies, both researchers classified all papers individualy before comparing the results. As before, in cases where there was disagreement, the issue was discussed until consensus was reached.

Finally, there is a potential bias in that both authors have written papers that were included in the review. Where only one author had participated in the primary study, the other author decided whether or not to include it.

## 6. Conclusion

This systematic review has addressed the following research questions. 1) What are the major knowledge management concepts that have been investigated in software engineering? 2) What are the major findings on knowledge management in software engineering? 3) What research methods have been used within the area so far?

For the first research question, our main findings are:
- The majority of studies of knowledge management in software engineering relate to technocratic and behavioural aspects of knowledge management.
- The studies that report on concepts within the fields of technocratic and behavioural aspects have very little overlap.
- There are few studies relating to economic, spatial and cartographic approaches to knowledge management.

For the second research question, we found that:
- As for the concepts, the findings are also divided and have very little overlap.
- The major finding, which is repeated over several papers and across several schools is the need to not focus exclusively on explicit knowledge but also on tacit knowledge.

For the third research question, we found that:
- The majority of reports of applications of knowledge management in the software engineering industry are reports of lessons learned, not scientific studies.
- Of the reports categorized as empirical studies, more than half of the reports are case studies.

- Our search returned field studies, action research, ethnographic studies, and one laboratory experiment.

The main implication for research is to focus more on the organizational school, while we believe practitioners should focus also on activities to manage tacit knowledge when working on knowledge management initiatives.

**Acknowledgement**

We are grateful to Reidar Conradi at the Department of Computer and Information Science, Norwegian University of Science and Technology, and Tore Dybå at SINTEF ICT for comments on an earlier version of this article. We would also like to thank Chris Wright for proofreading and useful comments. This work was partially funded by the Research Council of Norway through the project Evidence-Based Software Engineering (181685/I30).

**References**

1. S.A. Ajila and Z. Sun, Knowledge management: Impact of knowledge delivery factors on software product development efficiency, Proceedings of the IEEE International Conference on Information Reuse and Integration, Las Vegas, NV, United States, 2004, pp. 320-325.
2. A.J. Al-Shehab, R.T. Hughes, and G. Winstanley, Facilitating Organisational Learning Through Causal Mapping, Proceedings of the 7th International Workshop on Learning Software Organizations, Springer Verlag, Kaiserslautern, Germany, 2005, pp. 145-154.
3. M. Alavi and D.E. Leidner. Review: Knowledge Management and Knowledge Management Systems: Conceptual Foundations and Research Issues. MIS Quarterly. 25(1) (2001) 107-136
4. N. Angkasaputra, D. Pfahl, E. Ras, and S. Trapp, The Collaborative Learning Methodology CORONET-Train: Implementation and Guidance, Proceedings of the 4th International Workshop on Learning Software Organizations, Springer Verlag, Chicago, IL, USA, 2002, pp. 13-24.
5. J. Arent and J. Norbjerg, Software process improvement as organizational knowledge creation: a multiple case analysis, Proceedings of the Hawaii International Conference on System Sciences, Maui, USA, 2000, pp. 105.
6. J. Arent, J. Nørbjerg, and M.H. Pedersen, Creating Organizational Knowledge in Software Process Improvement, Proceedings of the 2nd Workshop on Learning Software Organizations, Oulu, Finland, 2000, pp. 81-92.
7. L. Argote, B. McEvily, and R. Reagans. Managing Knowledge in Organizations: An Integrative Framework and Review of Emerging Themes. Management Science. 49(4) (2003) 571-582
8. C. Argyris, Overcoming Organizational Defences: Facilitating Organizational Learning, Prentice Hall, 1990

9. C. Argyris and D.A. Schön, Organizational Learning II: Theory, Method and Practise, Organization Development Series, Addison Wesley, 1996
10. A. Aurum, R. Jeffrey, C. Wohlin, and M. Handzic, Managing Software Engineering Knowledge, Springer-Verlag, 2003
11. M.d.O. Barros, C.M.L. Werner, and G.H. Travassos. Supporting risks in software project management. Journal of Systems and Software. 70(1-2) (2004) 21-35
12. V.R. Basili, G. Caldiera, F. McGarry, R. Pajerski, and G. Page, The Software Engineering Laboratory - An operational software experience factory, Proceedings of the 14th International Conference on Software Engineering, 1992, pp. 370-381.
13. V.R. Basili, G. Caldiera, and H.D. Rombach, The Experience Factory, in: J.J. Marciniak (Eds.), Encyclopedia of Software Engineering, 1, John Wiley, 1994, pp. 469-476.
14. R. Baskerville and J. Pries-Heje. Knowledge capability and maturity in software management (1999
15. P.-H.J. Baskerville Richard. Knowledge capability and maturity in software management (1999
16. A. Birk, A Knowledge Management Infrastructure for Systematic Improvement in Software Engineering, Dr. Ing thesis, University of Kaiserslautern, Department of Informatics, 2000
17. F.O. Bjornson and T. Dingsoyr, A study of a mentoring program for knowledge transfer in a small software consultancy company, Proceedings of, Springer Verlag, Heidelberg, D-69121, Germany, Oulu, Finland, 2005, pp. 245-256.
18. F.O. Bjørnson, Knowledge Management in Software Process Improvement, PhD Thesis, Norwegian University of Science and Technology, Department of Computer and Information Science, 2007
19. F.O. Bjørnson and T. Dingsoyr, A study of a mentoring program for knowledge transfer in a small software consultancy company, in: Lecture Notes in Computer Science 3547, Springer Verlag, Heidelberg, 2005, pp. 245-256.
20. F.O. Bjørnson and T. Stålhane, Harvesting Knowledge through a Method Framework in an Electronic Process Guide, Proceedings of the 7th International Workshop on Learning Software Organizations, Springer Verlag, Kaiserslautern, Germany, 2005, pp. 86-90.
21. P. Brössler, Knowledge Management at a Software Engineering Company - An Experience Report, Proceedings of the 1st Workshop on Learning Software Organizations, Kaiserslautern, Germany, 1999, pp. 77-86.
22. A.F. Buono and F. Poulfelt, Challenges and Issues in Knowledge Management, Information Age Publishing, 2005
23. B. Chatters, Implementing an experience factory: maintenance and evolution of the software and systems development process, Proceedings of, 1999, pp. 146-151.
24. C.M. Chewar and D.S. McCrickard, Links for a human-centered science of design: Integrated design knowledge environments for a software development process, Proceedings of the Hawaii International Conference on System Sciences, Big Island, HI, United States, 2005, pp. 256.
25. C.W. Choo, The Knowing Organization: How Organizations Use Information to Construct Meaning, Create Knowledge, and Make Decisions, Oxford University Press, 1998

26. P.H. Christensen, The Wonderful World of Knowledge Management, in: A.F. Buono and F. Poulfelt (Eds.), Challenges and Issues in Knowledge Management, Information Age Publishing, 2005, pp. 337-364.
27. R. Conradi and T. Dybå, An empirical study on the utility of formal routines to transfer knowledge and experience, Proceedings of the ACM SIGSOFT Symposium on the Foundations of Software Engineering, Association for Computing Machinery, Vienna, Austria, 2001, pp. 268-276.
28. T.H. Davenport and L. Prusak, Working Knowledge: How Organizations Manage What They Know, Harvard Business School Press, 1998
29. R. De Almeida Falbo, L.S.M. Borges, and F.F.R. Valente, Using knowledge management to improve software process performance in a CMM level 3 organization, Proceedings of the Fourth International Conference on Quality Software, IEEE Computer Society, Braunschweig, Germany, 2004, pp. 162-169.
30. K.C. Desouza. Facilitating tacit knowledge exchange. Communications of the ACM. 46(6) (2003) 85-88
31. K.C. Desouza, Y. Awazu, and Y. Wan. Factors governing the consumption of explicit knowledge. Journal of the American Society for Information Science and Technology. 57(1) (2006) 36-43
32. K.C. Desouza, T. Dingsoyr, and Y. Awazu. Experiences with conducting project postmortems: Reports versus stories. Software Process Improvement and Practice. 10(2) (2005) 203-215
33. M. Dierkes, A. Berthoin Antal, J. Child, and I. Nonaka, Handbook of Organizational Learning and Knowledge, Oxford University Press, 2001
34. T. Dingsoyr, H.K. Djarraya, and E. Royrvik. Practical knowledge management tool use in a software consulting company. Communications of the ACM. 48(12) (2005) 96-100
35. T. Dingsoyr and E. Royrvik, An empirical study of an informal knowledge repository in a medium-sized software consulting company, Proceedings of the International Conference on Software Engineering, Portland, OR, United States, 2003, pp. 84-92.
36. T. Dingsøyr, Knowledge Management in Medium-Sized Software Consulting Companies, Dr. ing. thesis, Norwegian University of Science and Technology, Department of Computer and Information Science, 2002
37. T. Dingsøyr and R. Conradi. A survey of case studies of the use of knowledge management in software engineering. International Journal of Software Engineering and Knowledge Engineering. 12(4) (2002) 391-414
38. T. Dingsøyr and G.K. Hanssen, Extending Agile Methods: Postmortem Reviews as Extended Feedback, Proceedings of the 4th International Workshop on Learning Software Organizations, Springer Verlag, Chicago, IL, USA, 2002, pp. 4-12.
39. T. Dingsøyr and N.B. Moe. The Impact of Employee Participation on the Use of an Electronic Process Guide: A Longitudinal Case Study. IEEE Transactions on Software Engineering. To appear (2008

40. H.D. Doran, Agile Knowledge Management in Practice, Proceedings of the 6th International Workshop on Learning Software Organizations, Springer Verlag, Banff, Canada, 2004, pp. 137-143.

41. T. Dybå, Enabling Software Process Improvement: An Investigation on the Importance of Organizational Issues, Dr. ing thesis, Norwegian University of Science and Technology, Department of Computer and Information Science, 2001

42. T. Dybå, T. Dingsøyr, and G.K. Hanssen, Applying Systematic Reviews to Diverse Study Types: An Experience Report, Proceedings of the ESEM, Madrid, Spain, 2007,

43. T. Dybå, B.A. Kitchenham, and M. Jørgensen. Evidence-Based Software Engineering for Practitioners. IEEE Software. 22(1) (2005) 58-65

44. M. Earl. Knowledge Management Strategies: Towards a Taxonomy. Journal of Management Information Systems. 18(1) (2001) 215-233

45. M. Easterby-Smith and M.A. Lyles, The Blackwell handbook of organizational learning and knowledge management, Blackwell Publishing, 2003

46. C. Ebert, J. De Man, and F. Schelenz, e-R&D: Effectively Managing and Using R&D Knowledge, in: A. Aurum, et al. (Eds.), Managing Software Engineering Knowledge, Springer-Verlag, 2003, pp. 339-359.

47. J.S. Edwards, Managing Software Engineers and Their Knowledge, in: A. Aurum, et al. (Eds.), Managing Software Engineering Knowledge, Springer-Verlag, Berlin, 2003, pp. 5-27.

48. F.F. Fajtak, Kick-off Workshops and Project Retrospectives: A good learning software organization practice, Proceedings of the 7th International Workshop on Learning Software Organizations, Springer Verlag, Kaiserslautern, Germany, 2005, pp. 76-81.

49. P. Feher and A. Gabor. The role of knowledge management supporters in software development companies. Software Process Improvement and Practice. 11(3) (2006) 251-260

50. R.L. Feldmann and K.-D. Althoff, On the Status of Learning Software Organisations in the Year 2001, Proceedings of the Learning Software Organizations Workshop, Springer Verlag, Kaiserslautern, Germany, 2001, pp. 2-6.

51. H. Folkestad, E. Pilskog, and B. Tessem, Effects of Software Process in Organization Development – A Case Study, Proceedings of the 6th International Workshop on Learning Software Organizations, Springer Verlag, Banff, Canada, 2004, pp. 153-164.

52. R.L. Glass, V. Ramesh, and V. Iris. An Analysis of Research in Computing Disciplines. Communications of the ACM. 47(6) (2004) 89-94

53. G. Grabher and O. Ibert. Bad company? The ambiguity of personal knowledge networks. Journal of Economic Geography. 6(3) (2006) 251-271

54. B.H. Hansen and K. Kautz, Knowledge mapping: A technique for identifying knowledge flows in software organisations, in: Lecture Notes in Computer Science 3281, 2004, pp. 126-137.

55. M.T. Hansen, N. Nohria, and T. Tierney. What is your strategy for managing knowledge? Harvard Business Review. 77(2) (1999) 106 - 116

56. D. Hislop. Mission impossible? Communicating and sharing knowledge via information technology. Journal of Information Technology. 17 (2002) 165-177

57. F. Houdek and C. Bunse, Transferring Experience - A practical Approach and its Application on Software Inspections, Proceedings of the 1st Workshop on Learning Software Organizations, Kaiserslautern, Germany, 1999, pp. 59-68.
58. F. Houdek, K. Schneider, and E. Wieser, Establishing Experience Factories at Daimler-Benz. An Experience Report, Proceedings of the 20th International Conference on Software Engineering, Kyoto, Japan, 1998, pp. 443-447.
59. P. Jalote, Knowledge Infrastructure for Project Management, in: A. Aurum, et al. (Eds.), Managing Software Engineering Knowledge, Springer-Verlag, 2003, pp. 361-375.
60. C. Johannson, P. Hall, and M. Coquard, Talk to Paula and Peter - They are Experienced, Proceedings of the 1st Workshop on Learning Software Organizations, Kaiserslautern, Germany, 1999, pp. 69-76.
61. T. Kahkonen, Agile methods for large organizations - Building communities of practice, Proceedings of the Agile Development Conference, IEEE Computer Society, Salt Lake City, UT, United States, 2004, pp. 2-10.
62. A. Kankanhalli, B.C.Y. Tan, and K.-K. Wei. Contributing Knowledge to Electronic Knowledge Repositories: An Empirical Investigation. MIS Quarterly. 29 (2005) 113-143
63. K. Kautz and K. Thaysen. Knowledge, learning and IT support in a small software company. Journal of Knowledge Management. 5(4) (2001) 349-357
64. P. Kess and H. Haapasalo. Knowledge creation through a project review process in software production. International Journal of Production Economics. 80(1) (2002) 49-55
65. P. Kettunen. Managing embedded software project team knowledge. IEE Software. 150(6) (2003) 359-366
66. B.A. Kitchenham, *Procedures for Performing Systematic Reviews*, in *Technical Report TR/SE-0401*. 2004, Keele University.
67. B.A. Kitchenham, T. Dybå, and M. Jørgensen, Evidence-Based Software Engineering, Proceedings of the International Conference on Software Engineering, 2004, pp. 273-281.
68. S. Koenig, Integrated process and knowledge management for product definition, development and delivery, Proceedings of, 2003, pp. 133-141.
69. A. Koennecker, J. Ross, and L. Graham, Lessons Learned From the Failure of an Experience Base Initiative Using a Bottom-Up Development Paradigm, Proceedings of the 24th Annual NASA Software Engineering Workshop, Washington, USA, 1999,
70. D. Kolb, Experiental Learning: Experience as the Source of Learning and Development, Prentice Hall, 1984
71. D. Kolb, Management and the learning process, in: K. Starkey (Eds.), How Organizations Learn, Thomson Business Press, London, 1996, pp. 270-287.
72. F. Kurniawati and R. Jeffery, The long-term effects of an EPG/ER in a small software organisation, Proceedings of the Australian Software Engineering Conference, Melbourne, Vic., Australia, 2004, pp. 128-136.
73. D. Landes, K. Schneider, and F. Houdek. Organizational learning and experience documentation in industrial software projects. International Journal of Human Computer Studies. 51(3) (1999) 643-661

74. S.-h. Liao. Knowledge management technologies and applications - literature review from 1995 to 2002. Expert Systems with Applications. 25 (2003) 155-164
75. J. Liebowitz. A look at NASA Goddard space flight center's knowledge management initiatives. IEEE Software. 19(3) (2002) 40-42
76. M. Lindvall, P. Costa, and R. Tesoriero, Lessons Learned about Structuring and Describing Experience for Three Experience Bases, Proceedings of the 3rd International Workshop on Learning Software Organizations, Springer Verlag, Kaiserslautern, Germany, 2001, pp. 106-119.
77. M. Lindvall and I. Rus. Knowledge Management in Software Engineering. IEEE Software. 19(3) (2002) 26 - 38
78. M. Lindvall and I. Rus, Knowledge Management for Software Organizations, in: A. Aybüke, et al. (Eds.), Managing Software Engineering Knowledge, Springer Verlag Berlin, 2003, pp. 73-94.
79. M. Lindvall and I. Rus, Lessons Learned from Implementing Experience Factories in Software Organizations, Proceedings of the 5th International Workshop on Learning Software Organizations, Bonner Köllen Verlag, Luzern, Switzerland, 2003, pp. 59-64.
80. M. Lindvall, I. Rus, R. Jammalamadaka, and R. Thakker, *Software Tools for Knowledge Management*, in *tech. report*. 2001, DoD Data Analysis Center for Software, Rome, N.Y.
81. K. Lyytinen and D. Robey. Learning failure in information systems development Information Systems Journal. 9(2) (1999) 85-101
82. M. Markkula, Knowledge Management in Software Engineering Projects, Proceedings of the International Conference on Software Engineering and Knowledge Engineering, Kaiserslautern, Germany, 1999, pp. 20-27.
83. N. Martin-Vivaldi, P. Collier, and S. Kipling, Peer Performance Coaching: Accelerating Organizational Improvement through Individual Improvement, Proceedings of the 2nd Workshop on Learning Software Organizations, Oulu, Finland, 2000, pp. 103-112.
84. L. Mathiassen and L. Vogelsang, The role of networks and networking in bringing software methods to practice, Proceedings of the Hawaii International Conference on System Sciences, Big Island, HI, United States, 2005, pp. 256.
85. R. McDermott. Why Information Technology Inspired But Cannot Deliver Knowledge Management. California Management Review. 41 (1999) 103-117
86. G. Melnik and F. Maurer, Direct verbal communication as a catalyst of agile knowledge sharing, Proceedings of the Agile Development Conference, Salt Lake City, UT, United States, 2004, pp. 21-31.
87. K. Mohan and B. Ramesh, Managing variability with traceability in product and service families, Proceedings of, 2002, pp. 1309-1317.
88. S. Nerur and V. Balijepally. Theoretical Reflections on Agile Development Methodologies. Communications of the ACM. 50 (2007) 79-83
89. E. Niemela, J. Kalaoja, and P. Lago. Toward an architectural knowledge base for wireless service engineering. Ieee Transactions on Software Engineering. 31(5) (2005) 361-379
90. I. Nonaka and H. Takeuchi, The Knowledge-Creating Company, Oxford University Press, 1995

91. J. Nørbjerg, T. Elisberg, and J. Pries-Heje, Experiences from using knowledge networks for sustaining Software Process Improvement, Proceedings of the 8th International Workshop on Learning Software Organizations, Rio de Janeiro, Brazil, 2006, pp. 9-17.

92. T.J. Ostrand and E.J. Weyuker, A Learning Environment for Software Testers at AT&T, Proceedings of the 2nd Workshop on Learning Software Organizations, Oulu, Finland, 2000, pp. 47-54.

93. M.C. Paulk, C.V. Weber, and B. Curtis, The Capability Maturity Model: Guidelines for Improving the Software Process, Addison-Wesley, Reading, MA, USA, 1995

94. M. Polanyi, The Tacit Dimension, Doubleday, 1967

95. S. Ramasubramanian and G. Jagadeesan. Knowledge management at infosys. Ieee Software. 19(3) (2002) 53-+

96. E. Ras, G. Avram, P. Waterson, and S. Weibelzahl. Using weblogs for knowledge sharing and learning in information spaces. Journal of Universal Computer Science. 11(3) (2005) 394-409

97. T. Ravichandran and A. Rai. Structural analysis of the impact of knowledge creation and knowledge embedding on software process capability. Ieee Transactions on Engineering Management. 50(3) (2003) 270-284

98. O.M. Rodriguez, A.I. Martinez, A. Vizcaino, J. Favela, and M. Piattini, Identifying knowledge management needs in software maintenance groups: A qualitative approach, Proceedings of the Fifth Mexican International Conference in Computer Science, Colima, Mexico, 2004, pp. 72-79.

99. T.R. Roth-Berghofer, Learning from HOMER, a Case-Based Help Desk Support System, Proceedings of the 6th International Workshop on Learning Software Organizations, Springer Verlag, Banff, Canada, 2004, pp. 88-97.

100. I. Rus, M. Lindvall, and S.S. Sinha, *Knowledge Management in Software Engineering*, in *tech. report*. 2001, DoD Data Analysis Center for Software, Rome.

101. O. Salo, Systematical Validation of Learning in Agile Software Development Environment, Proceedings of the 7th International Workshop on Learning Software Organizations, Springer Verlag, Kaiserslautern, Germany, 2005, pp. 106-110.

102. K. Schneider. What to expect from software experience exploitation. Journal of Universal Computer Science. 8(6) (2002) 570-580

103. K. Schneider, J.-P. Von Hunnius, and V.R. Basili. Experience in implementing a learning software organization. IEEE Software. 19(3) (2002) 46-49

104. J.-P.v.H. Schneider Kurt. Experience reports: process and tools: Effective experience repositories for software engineering (2003

105. U. Schultze and D.E. Leidner. Studying Knowledge Management in Information Systems Research: Discourses and Theoretical Assumptions. MIS Quarterly 26 (2002) 213-242

106. D.G. Schwartz, Encyclopedia of Knowledge Management, Idea Group Reference, 2006

107. L. Scott and T. Stålhane, Experience Repositories and the Postmortem, Proceedings of the 5th International Workshop on Learning Software Organizations, Bonner Köllen Verlag, Luzern, Switzerland, 2003, pp. 79-82.

108. J. Segal, Organisational learning and software process improvement: a case study, Proceedings of the 3rd International Workshop on Learning Software Organizations, Springer Verlag, Kaiserslautern, Germany, 2001, pp. 68-82.
109. P.M. Senge, The Fifth Discipline: The Art & Practise of The Learning Organisation, Century Business, 1990
110. H. Sharp and H. Robinson. An ethnographic study of XP practice. Empirical Software Engineering. 9 (2004) 353-375
111. D. Skuce. Knowledge Management in Software-Design - a Tool and a Trial. Software Engineering Journal. 10(5) (1995) 183-193
112. R. Stata, Organizational learning: The key to management innovation, in: K. Starkey (Eds.), How organizations learn, Thomson Business Press, London, 1996, pp. 316-334.
113. K.E. Sveiby, The New Organizational Wealth: Managing and Measuring Knowledge-Based Assets, Berret-Koehler Pub, 1997
114. J. Swan, H. Scarbrough, and J. Preston, Knowledge Management - The Next Fad to Forget People?, Proceedings of the 7th European Conference on Information Systems, 1999, pp. 668-678.
115. A.H. Torres, N. Anquetil, and K. Oliveira, Pro-active dissemination of Knowledge with Learning Histories, Proceedings of the 8th International Workshop on Learning Software Organizations, Rio de Janeiro, Brazil, 2006, pp. 19-27.
116. R. Trittmann, The organic and the mechanistic form of managing knowledge in software development, Proceedings of the 3rd International Workshop on Learning Software Organizations, Springer Verlag, Kaiserslautern, Germany, 2001, pp. 22-26.
117. J.-W. van Aalst, Knowledge Management in Courseware Development, PhD thesis, Technical University Delft, 2001
118. C.G.v. Wangenheim, S. Weber, J.C.R. Hauck, and G. Trentin. Experiences on establishing software processes in small companies. Information and Software Technology. 48(9) (2006) 890-900
119. J. Ward and A. Aurum, Knowledge management in software engineering - Describing the process, Proceedings of the Australian Software Engineering Conference, Melbourne, Vic., Australia, 2004, pp. 137-146.
120. E. Wenger, Communities of practice : learning, meaning and identity, Cambridge University Press, 1998
121. K.P. Yglesias, IBM's reuse programs: Knowledge management and software reuse, Proceedings of the International Conference on Software Reuse, 1998, pp. 156-164.

# Appendix

**Table 11: Categorized articles, extended**

|     | Systems | Cartographic | Engineering | Economic | Organizational | Spatial | Strategic |
|---|---|---|---|---|---|---|---|
| Emp | [11, 20, 24, 35, 72, 111] | [34] | [2, 6, 15, 17, 27, 32, 51, 54, 86, 101, 108, 118] |  | [53, 84, 91] |  | [5, 49, 116] |
| LL | [4, 12, 23, 29, 57-59, 68, 69, 73, 76, 79, 82, 87, 89, 96, 99, 102, 104, 107] |  | [4, 38, 48, 64, 65, 83, 98, 107, 115] |  | [60, 61] | [30] | [21, 40, 46, 63, 75, 92, 95, 103, 121] |

**Table 12: Overview of research methods, extended**

| Research Method | KM/SE |
|---|---|
| Action Research | [5, 17, 20] |
| Case study | [2, 6, 15, 24, 32, 51, 53, 72, 84, 91, 101, 108, 111, 118] |
| Ethnography | [34, 35] |
| Laboratory Experiment | [86] |
| Field Study | [11, 27, 49, 54, 116] |